\documentclass[
twocolumn,
aps,floatfix,nofootinbib]{revtex4}
\usepackage{latexsym}
\usepackage{mathptm}
\usepackage{amsmath}
\usepackage{amssymb}
\usepackage{graphicx}
\def\beq{\begin{equation}}
\def\eeq{\end{equation}}
\def\beqn{\begin{eqnarray}}
\def\eeqn{\end{eqnarray}}
\newcommand{\f}{\begin{equation}}
\newcommand{\ff}{\end{equation}}
\newcommand{\be}{\begin{equation}}
\newcommand{\ee}{\end{equation}}
\newcommand{\barray}{\begin{array}}
\newcommand{\earray}{\end{array}}
\newcommand{\bea}{\begin{eqnarray}}
\newcommand{\eea}{\end{eqnarray}}

\begin{document}
\title{General relativity as the equation of state of spin foam }
\author{ Lee Smolin  \\
Perimeter Institute for Theoretical Physics, \\ 31 Caroline Street North, Waterloo, Ontario N2J 2Y5, Canada}

\date{\today}

\begin{abstract}

Building on recent significant results of Frodden, Ghosh and Perez (FGP) and Bianchi, I present a quantum version of Jacobson's argument that the Einstein
equations emerge as the equation of state of a quantum gravitational system.  I give three criteria a quantum theory of gravity must satisfy if it is
to allow Jacobson's argument to be run.  I then show that the results of FGP and Bianchi provide evidence that loop quantum gravity satisfies two of these criteria and argue that the third should also be satisfied in loop quantum gravity.  I also show that the energy defined by
FGP is the canonical energy associated with the boundary term of the Holst action.  

\end{abstract}

\maketitle

\tableofcontents

\section{Introduction}

In 1995 Ted Jacobson gave a highly influential argument that the Einstein equations arise from the thermodynamics of a deeper theory as the equation of state\cite{Ted1}.  Here I give a quantum mechanical version of that argument and show that it can be run in loop quantum gravity when the dynamics are given in terms of a spin foam model.  This is possible because of recent results of Frodden, Ghosh and Perez\cite{FGP} (FGP) and Bianchi\cite{Eugenio1} on the thermodynamics of quantum black holes in loop quantum gravity.  These works exploit, in the context of loop quantum gravity, the analysis of the thermodynamics of black holes as seen from an observer hovering a short distance above the black  hole horizon\cite{CT,MP,JP}.
This gives a new argument that the classical Einstein equations are the correct classical dynamics to govern an emergent classical space-time description of spin foam models.  

To make it clear what is assumed and what is shown here I give a brief sketch of the main argument.  

\begin{itemize}

\item{} 
I assume that there is an underlying dynamical quantum geometry which has a background independent description in the language of loop quantum gravity.   In order to utilize the results of FGP and Bianchi, the quantum dynamics is given according to recent work in spin foam 
models\cite{reviewSF}.

\item{} 
I {\it assume} that there are large regions of space-time which admit a course grained description in terms of a classical space-time manifold,
$\cal M$, on which there is defined a classical metric $g_{ab}$.  There are also assumed to be matter fields on the quantum space-time whose emergent course grained description is in terms of a conserved energy-momentum tensor $T_{ab}$ defined on $\cal M$.  
Both $g_{ab}$ and $T_{ab}$ are assumed to be slowly varying on the Planck scale.  What is not assumed is that
$g_{ab}$ and $T_{ab}$ are related by the Einstein equations, this is what is to be shown.

\item{} 
I propose a new context to relate the classical and quantum description of space-time, which is suggested by the work of \cite{FGP}.  This is inspired by an idea used in particle physics which is the the infinite-momentum frame\cite{IMF}.  The idea, as developed by Bjorken,  was that by transforming to an arbitrarily boosted frame of reference all physical momenta are sent to infinity, which reveals the ultraviolet properties of matter.  This was successful in revealing the parton structure of hadrons.  I suggest a twist on this idea may be used to reveal the microscopic structure of quantum geometry. In this case the idea is that a highly boosted accelerated observer can probe the quantum geometry.  

The Unruh effect can be interpreted as an indication that accelerated observers probe the microscopic structure of quantum fields\cite{Unruh}.  The exciting idea pioneered by  FGP\cite{FGP} and confirmed by results of Bianchi\cite{Eugenio1} is that accelerated observers can also probe the microscopic quantum geometry.

\end{itemize}

To implement this idea we consider a region of $\cal M$ of size $L$, small compared to the maximal radius of curvature. In this region we imagine a uniformly accelerating observer, with four acceleration $a$ such that $l=a^{-1} << L$.  The observer's motion creates for her an horizon, a distance
$l$ behind her.  By extending perpendicularly to the observer's world line we construct a family of parallel accelerated observers, whose worldlines make a three dimensional timelike surface, $\cal S$.  Their horizons then form two light like sheets, $H^\pm$ that intersect at the space like two surface, $H$.  

\begin{figure}[h!]
\begin{center}
\includegraphics[width=.5 \textwidth]{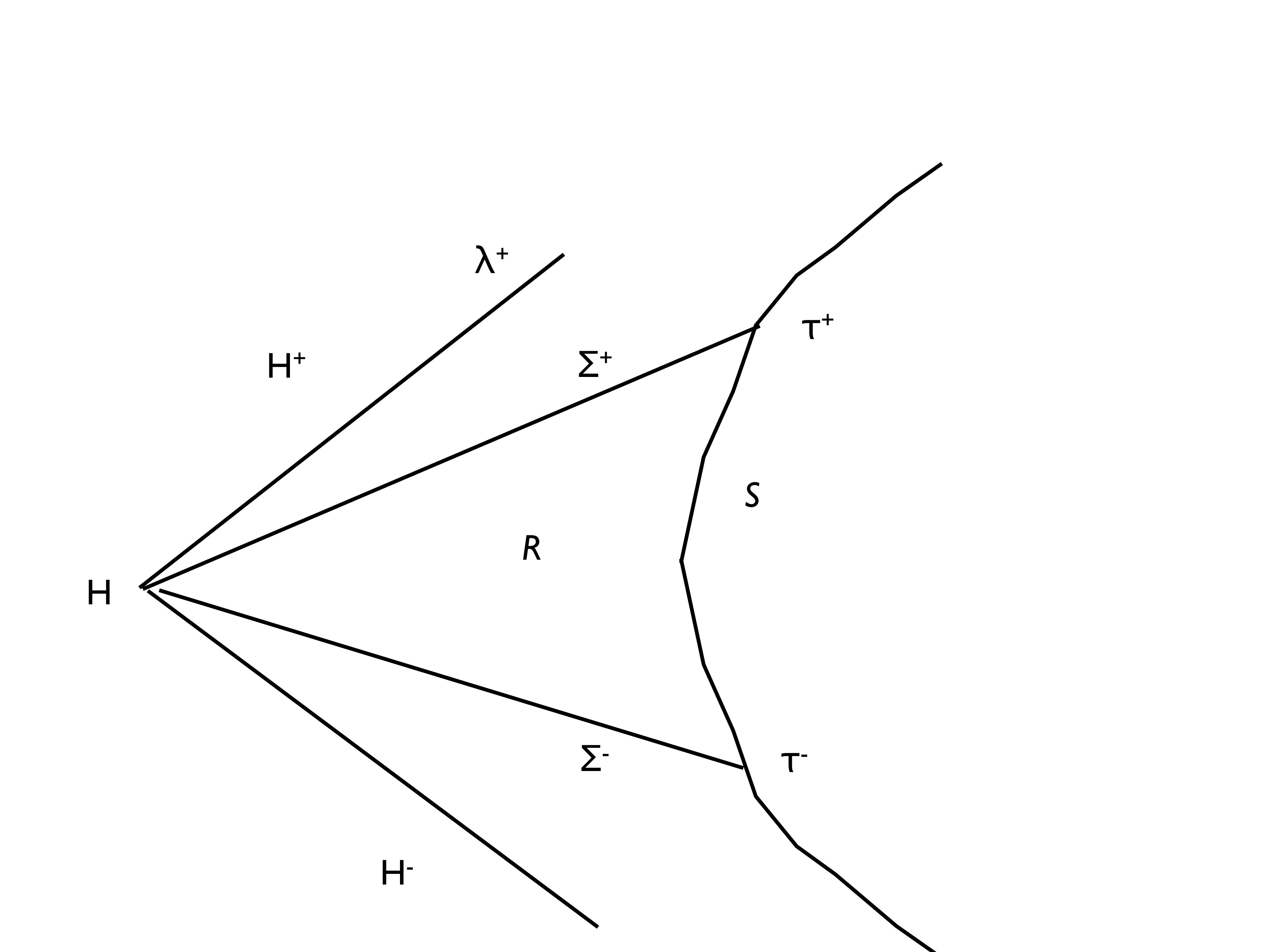}
\end{center}
\caption{A quantum near horizon region, $\cal R$ defined by a family of accelerated observers on $\cal S$.}
\end{figure}

The observer carries a clock which measures a time $\tau$.  To zeroth order in $\frac{1}{L}$, the geometry of the region between the observer's world line and the horizon can be approximated by Rindler space-time, as is shown in Figure 1.

We define a subregion of $\cal M$ denoted $\cal R$ which is a finite piece of the near horizon region observed by the observers on $\cal S$.  
Pick two times on the observer's world line, $\tau_-$ and $\tau_+$. Each has a three dimensional spacelike surface of simultaneity, $\Sigma^\pm$ that intersect at $H$.  

The construction is completed by the erection of a timelike wall, $W, $ at constant values of the transverse Rindler coordinates, so that the cross-sectional area enclosed is finite.  We can then define 
$\cal R$ to consist of the region bounded by the surfaces, $\cal S$, $\Sigma^\pm$ , $W$ and $H$.   


We know from the Unruh effect that when quantum field theory on the Rindler geometry is considered,  ${\cal R}$ is a thermodynamic system which, as measured by observers on $\cal S$, has a finite temperature. But that is just an approximation.  The full description of $\cal R$ is as a dynamical quantum geometry.  We can ask whether $\cal R$ can still be considered a thermodynamic system when studied with the tools of quantum gravity. The results of FGP and Bianchi tell us it is.  


The idea then is to give a deeper description of the  region $\cal R$ by replacing its coarse grained description in terms of a classical metric and fields with a region of quantum space-time geometry.  This quantum description of $\cal R$ can be called a quantum near horizon geometry.  The states of this quantum description live in a Hilbert space, ${\cal H}_{\cal R}$ which has a precise construction in loop quantum gravity.  The quantum geometry dynamics are observed by the family of accelerating observers who make up its timelike boundary, $\cal S$.  Those observers can gain information about the quantum geometry in $\cal R$ by measuring suitable operators defined on their worldlines which make up its timelike boundary.  


We can note that spin foam models are formulated precisely to investigate regions of quantum space-time with classical boundaries 
such as $\cal R$ \cite{reviewSF}.  Following  standard ideas we join the quantum description of the space-time region, $\cal R$ to the classical description by equating the pull back of the classical metric on its boundary 
$\partial {\cal R}$ with the expectation values of the corresponding quantum geometric operators.  

Note that the classical and quantum descriptions of $\cal R$ are just different levels of description of a single system.  Every macroscopic region of space-time has both a classical and a quantum description.  The argument to be given here relies on the fact that one can choose to describe an arbitrary near horizon region, created by the motion of an arbitrary accelerated observer, in either classical or quantum terms and that these two descriptions will be related as just specified.  


The quantum fields in the interior of $\cal R$ are then governed by quantum gravitational dynamics defined by constraints and a boundary Hamiltonian on $\cal S$. This defines an energy, which allows us to study the thermodynamics of the quantum system comprising the quantum geometry and matter within $\cal R$.  


The appropriate quantum Hamiltonian on the quantum system $\cal R$ has been defined by Bianchi\cite{Eugenio1} ,who shows that its expectation value agrees with the classical definition of the the energy seen by the accelerating observer, as defined by Frodden, Ghosh and Perez\cite{FGP}.   We show below the FGP energy is in fact the canonical hamiltonian for the near horizon region defined by evaluating the boundary terms of the Holst action on the accelerating timelike boundary $\cal S$.  


The first of the three properties to be imposed on the quantum description of the near horizon region can be considered a quantum version of the equivalence principle.  This holds that there exists a quantum state, $| R>$, such that the change in the expectation value of the FGP energy, due to a matter perturbation, is given by the flux of the energy momentum tensor through the timelike surface, $\cal S$.  This can be called the {\it Sciama} property.


Following Bianchi we presume that observers on $\cal S$ carry thermometers which couple to the quantum geometry degrees of freedom within 
$\cal R$.  The first criteria to be satisfied by a quantum theory of gravity is that there should be a state in the Hilbert space associated with $\cal R$
whose temperature so measured is the Unruh temperature\cite{Unruh},
\f
T=\frac{\hbar a}{2\pi c}
\ff
We will say a quantum theory of gravity in which this is true has the  {\it Unruh} property.   


We also define the entropy of the quantum near horizon geometry, $S$.  We say a quantum
theory of gravity has the {\it Bekenstein} property if the changes of this entropy obey
\f
\delta S = \frac{\delta A}{4 \hbar G}
\ff
where $A$ is the area of space like sections of $H^+$.  


We then consider perturbing the quantum system in $\cal R$ by making a small change to the energy momentum tensor, $\delta T_{ab}$ 
on $\cal S$.  We assume that the quantum gravitational system $\cal S$ satisfies the first law of thermodynamics\footnote{More properly, called the Clausius relation.}
\f
\delta S =\frac{\delta E}{T}
\ff



\begin{figure}[h!]
\begin{center}
\includegraphics[width=.5 \textwidth]{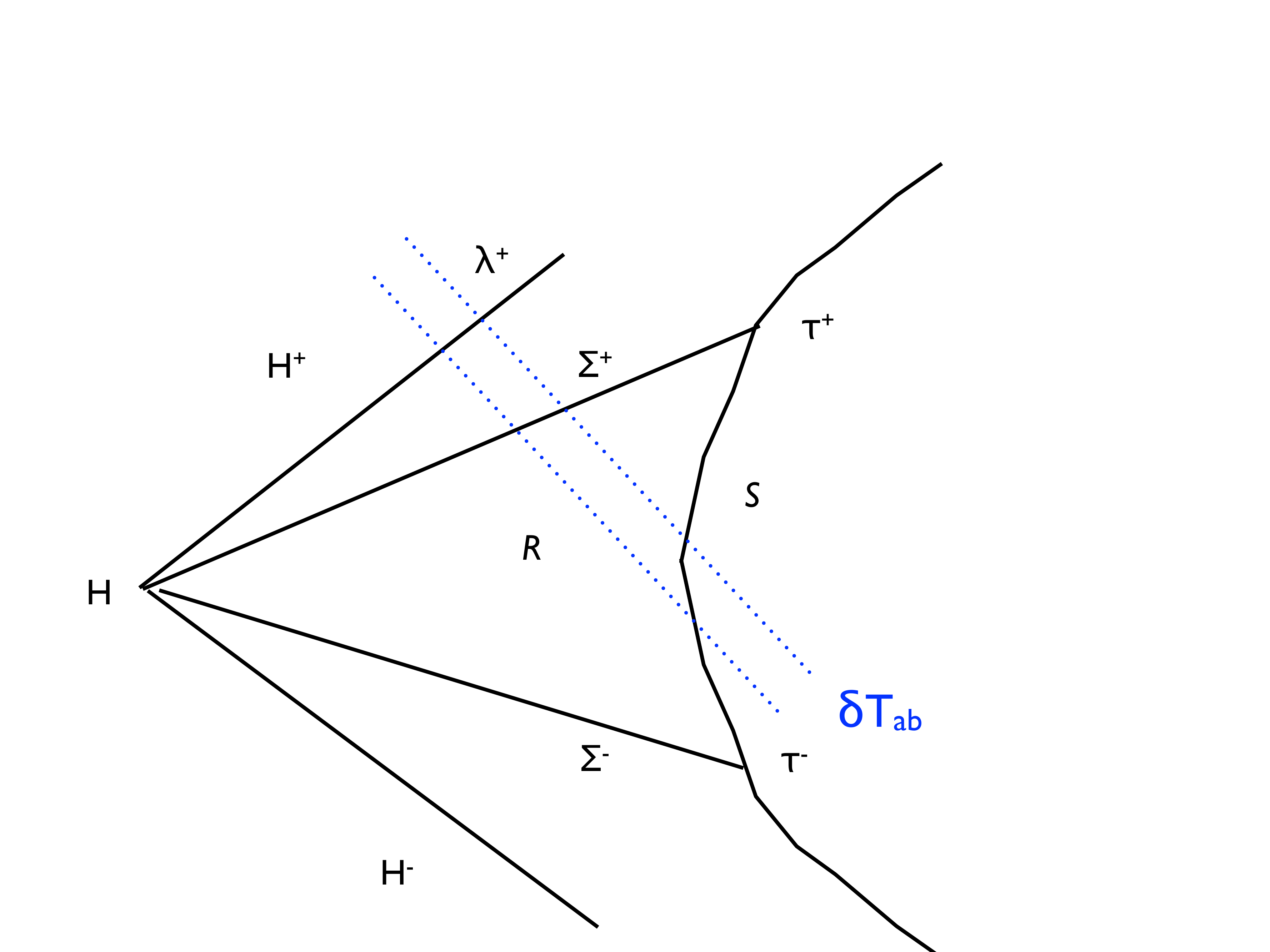}
\end{center}
\caption{A perturbation by a classical $\delta T_{ab}$.}
\end{figure}

Given these assumptions we will derive below two main results:

\begin{enumerate}

\item{} If a quantum theory of gravity is Sciama, Unruh and Bekenstein, then the classical metric and stress energy tensors will be related by the Einstein equations.  This is essentially a restatement of Jacobson's classic result\footnote{Very interesting related arguments have been given by Padmanabhan\cite{Paddy}.} in \cite{Ted1} and is the subject of section $III$.

\item{} The recent results of  FGP and Bianchi provide evidence that loop quantum gravity, with dynamics given in the spin foam formalism, is Unruh and Bekenstein. We also argue below that it will be Sciama because the quantum constraints are satisfied by spin foam amplitudes.   Hence,  there is good evidence that, if a spin foam model has an emergent description in terms of classical metric and energy momentum tensor, these are related by the Einstein equations. This is the subject of section $IV$.

\end{enumerate}

The presentations of these results depends on the construction of the classical and quantum dynamics of the near horizon region, which are presented in section $II$.  In section $V$ I discuss the leading order contributions in the spin foam sum over histories which contributes to the quantum dynamics of $\cal R$, before closing with brief conclusions.  

\section{The dynamics of the near horizon region}

\subsection{Classical dynamics of the near horizon geometry}

The classical near horizon geometry is defined by the Rindler metric,
\f
ds^2 = -\frac{a^2 \rho^2}{c^2} d\tau^2 + d\rho^2 + h_{ij} dx^i_\perp dx^j_\perp
\label{Rindler}
\ff
where $a$ is an acceleration.  $\chi^a =(\frac{\partial}{\partial \tau} )^a$ is the boost killing field whose norm is
\f
\chi^a \chi_a = -\frac{a^2 \rho^2}{c^2} 
\ff

We will be interested in regions, $\cal M$, of space time which are approximately Rindler, in that their geometry is defined by a metric,
\f
g_{ab}=g^R_{ab} + \delta g_{ab}
\ff
where $g^R_{ab}$ is the Rindler metric and $\delta g_{ab}$ is small and gives rise to a curvature whose radius of curvature is bounded below by
some large length $L$.  The  space-time region will also contain matter, represented by $\delta T_{ab}$, which we require to be small in the sense that
\f
|\delta T_{ab}| < \frac{1}{8\pi G L^2}
\label{bound}
\ff
so that,  if we succeed to show that the Einstein equations emerge,  the radius of curvature sourced by $\delta T_{ab}$ by those equations will be larger than $L$, so that the result we seek is consistent with the assumptions made during its derivation.

To apply thermodynamics to the near horizon geometry of a black hole we define the open system, $\cal R$,  defined above and shown in Figure 1.
The system $\cal R$ is defined by the interior of two space like surfaces, the timelike surface $\cal S$ and the timelike wall, $W$.

The outer, timelike boundary $\cal S$ is the surface $\rho_0 =\frac{c^2}{a}$.   The two space like boundaries are 
at $\tau=\tau_-$ and $\tau=\tau_+$, and are indicated by $\Sigma_\pm$, these meet at the horizon $\rho=0$, called $H$.

The total boundary of $\cal R$ is given by
\f
\partial {\cal R} = {\cal S} \cup \Sigma_+ \cup H \cup \Sigma_-  \cup W
\ff
To define the classical dynamics within $\cal R$ we allow the fields within $\cal R$ to vary subject to suitable boundary conditions
on $\partial {\cal R}$.   
We use an action (to be described below) from which we define a canonical
Hamltonian $H_{\cal R}$, conjugate to the time $\tau$ of the accelerating observers.  This Hamiltonian, $H$, will have a bulk term,
proportional to constraints, which we collectively denote, $\cal C$ and a boundary term, $H^{boundary}$ defined on $\cal S$.  

It is straightforward to derive the boundary term for the Hamiltonian.  We 
fix space-time to be Rindler at and outside of the boundaries $\partial {\cal R}$ and study the interior of $\cal R$ as a classical
dynamical system.  The dynamics is given by the Holst action with a boundary term
\begin{eqnarray}
S&= & S_{\cal R} + S_{\cal S}  \\
&=&  \int_{\cal R} \left \{ \frac{1}{8\pi G} \left ( (e \wedge e)^*_{IJ} +\frac{1}{\gamma} e_I \wedge e_J  \right )\wedge F^{IJ} + {\cal L}^{matter}
\right \} 
\nonumber \\
&&- \frac{1}{8\pi G} \int_{\cal S} \left ( (e \wedge e)^*_{IJ} +\frac{1}{\gamma} e_I \wedge e_J  \right ) \wedge A^{IJ}  \nonumber
\label{action}
\end{eqnarray}
This action is well defined if we fix the pullback of the $e_a$ on the outer boundary, $\cal S$, and let the other fields vary\cite{boundary}.  The philosophy is
that we are choosing the boundary conditions to model the influence of external fields on the near horizon dynamical geometry.    We also fix
the variation of the connection, $\delta A_a^{IJ} =0$ on the horizon two surface, $H$, and the wall, $W$, so no boundary contribution to the action or Hamiltonian arises
from there.

It is easy to evaluate the boundary term given that the geometry is assumed to be Rindler on the boundary.
We use the frame fields
\f
e^0= \frac{a \rho}{c} d\tau , \ \ \ e^\rho = d\rho , \ \ \ e^{i=1,2} = dx^i
\ff
The connection on solutions satisfies
\f
de^I= A^I_J \wedge e^J
\ff
The only non-vanishing component of $A^{IJ}$ is
\f
A^{0 \hat{\rho} } = \frac{a}{c} d\tau
\ff


It is straightforward now to evaluate 
\f
S_{\cal S} = \int d^2 x_\perp d\tau \tilde{\cal L}
\ff
where the boundary contribution to the Lagrangian is
\f
\tilde{\cal L}= \frac{1}{8\pi G} \sqrt{h} \frac{a}{c}
\ff
where $h$ is the determinant of the pull back of the metric into the boundary.
The Hamiltonian then has a boundary term
\f
H_{\cal S} = \int_{\partial \Sigma} \tilde{\cal L}= \frac{1}{8\pi G} A \frac{a}{c}
\label{HFGP}
\ff
which agrees with FGP.  Note that the Immirzi parameter does not appear.

We note that a related result has been derived by Carlip and Teitelboim\cite{CT} and Massar and Parentani\cite{MP}\footnote{as is reviewed by Jacobson and Parentani\cite{JP}.  This result has also been gotten in the connection variables by Bianchii and Wieland\cite{BW}.} who derive the boundary energy on the horizon, $H$, in the case that the dynamical region extends outwards to infinity.  In this case an energy arises from a contribution on the horizon which is proportional to the area on $H$, $A_H$.  

The Hamiltonian (\ref{HFGP}) already very remarkably contains the connection between energy and 
geometry that is encapsulated by the Einstein equations.  We will exploit this below.

\subsection{Quantum dynamics of the near horizon region}

We consider the quantum dynamics to the interior of the region $\cal R$.  We assume that outside of $\cal R$ there is a classical space-time metric,
$g_{ab}$ and energy momentum tensor, $T_{ab}$, which  are slowly varying on the Planck scale.  

The quantum dynamics of $\cal R$ is defined by the quantization of (\ref{action}) with the same boundary conditions.  
We  assume the quantum theory of gravity is coupled to matter degrees of freedom.  
This is described in the usual terms involving a Hilbert space of solutions to the quantum constraints including 
matter fields.  
\f
\hat{\cal C} |\Psi (\tau ) > =0
\label{constraints}
\ff
The states are functions of the time $\tau$ defined on the boundary but satisfy the quantum constraints for each time.
They evolve with respect to the boundary time,
\f
\imath \hbar \frac{d}{d \tau} |\Psi (\tau ) > =\hat{H}^{boundary} |\Psi (\tau ) > 
\ff

We couple the quantum dynamics in $\cal R$ to the classical dynamics in the external region by demanding that the pull backs 
of $g_{ab}$and $T_{ab}$  on
$\partial {\cal R}$  match the expectation values of the corresponding quantum operators on the space like potions of the  boundary.  They are
already matched on $\cal S$ by the common boundary conditions for the classical and quantum dynamics.  


The quantum theory should give us an amplitude for the following data:  
\begin{enumerate}

\item{} Classical data to be given on the  boundary $\partial {\cal R}$ including the pull backs of $g_{ab}$ and the  energy momentum tensor $T_{ab}$.

\item{} An incoming quantum state, $|\Psi_- >$ defined on $\Sigma_-$.

\item{} An outgoing quantum state, $|\Psi_+>$ defined on $\Sigma_+$.

\item{} The values of $\tau_\pm$ and $a$.

\end{enumerate}

Note that the classical data on $\partial {\cal R}$ constrains, but does not completely determine the choice of quantum states on
$\Sigma^\pm$.  

The external energy momentum tensor is assumed to be conserved which means that to leading order in $\frac{1}{L}$, 
\f
\int_{\partial {\cal R}} J_a N^a =0
\label{conserve2}
\ff  
where $J_a=T_{ab}\chi^b$ is the conserved energy current and $N^a$ is the unit normal to $\partial {\cal R}$. 

Given these data a quantum theory of gravity computes an amplitude
\f
Z(\Psi_\pm, T_{ab},\tau_\pm, a) = <\Psi_+ |  U (T_{ab},\tau_\pm, a) | \Psi_- >
\ff
where $U (T_{ab},\tau_\pm, a)$ is the quantum gravity evolution operator, which formally is given by
\f
U (T_{ab},\tau_\pm, a) = e^{\frac{\imath}{\hbar} \int_{\tau_-}^{\tau_+} d\tau \hat{H} }
\ff
The Hamiltonian generates evolution of the quantum state in $\tau$ which is the time coordinate of a boost killing field, as first shown in
\cite{CT,MP}.  
Correspondingly we can follow Bianchi to write,  
\f
H^{boundary} = \hbar a \hat{ K}^\rho
\label{boostH}
\ff 
where $\hat{K}^\rho$ is a generator of boosts acting in the hilbert space \cite{Eugenio1}.  

The amplitude $Z(\Psi_\pm, T_{ab},\tau_\pm, a)   $ may also be given by a path integral.  

We also will need there to be an area operator on spatial slices of the timelike boundary, denoted $\hat{A}_{\cal S} $ and an 
area operator on the inner boundary, $H$ denoted by $\hat{A}_{H} $. 


The matching of classical and quantum data implies  that initially the areas satisfy
\f
<\Psi_- | \hat{A}_{H} (\lambda = 0)|\Psi_- > =  < \Psi_- | \hat{A}_{\cal S} (\tau_-)|\Psi_- >
\label{initial}
\ff

\section{Jacobson's argument}

The goal of this section is to state criteria for a quantum theory of gravity such that the Einstein equations can be derived as an equation of state following Jacobson's argument. 
Following Jacobson\cite{Ted1},  we  work within a region of size $L$ less than the radius of curvature which can be considered approximately flat.  We pick a space like-two surface, $H$, which can be approximated as flat, in the sense that its past and future directed null normals congruences to one side have vanishing expansion and shear.   Those  incoming and outgoing null surfaces to the same side rooted at $H$ are denoted $H^\pm$.    The null geodesics comprising the outgoing surface $H^+$ are parameterized by an affine parameter, $\lambda$.  We then consider a family of accelerating observers with a constant acceleration, $a$, which have an horizon consisting of $H$ and $H^\pm$. The worldlines of these accelerating observers form a timelike surface $\cal S$.  The outside of $H, H^\pm$ can be parameterized by a Rindler coordinate system, as given in (\ref{Rindler}).

The quantum gravitational physics observed from $\cal S$ is described in terms of a quantum near horizon region which we have just described.

\subsection{Criteria}

I specify the three properties of the quantum theory of gravity necessary to run Jacobson's argument.  

\begin{itemize}

\item{} We call a quantum theory of gravity, {\it Sciama\footnote{Because of Dennis Sciama's search for a quantum version of the equivalence 
principle\cite{Sciama}.} } if it satisfies the following definition of a quantum equivalence principle:

There is a state $| R >$ called the "quantum Rindler state" such that the response to the matter perturbation $\delta T_{ab}$ of the expectation 
value of the boost energy (\ref{boostH}) is, to  leading order in $\frac{1}{L}$,
\f
\delta E = \delta <R| \hat{H}^{boundary} |R>     =\int_{\cal S} T_{ab} N^a \chi^b
\label{Ted1}
\ff

The justification for this is as follows: $H^{boundary}$ is a canonical Hamiltonian conjugate to the Rindler time, $\tau$, given, on physical states by the boundary hamiltonian. The Rindler state should correspond to flat Rindler space-time.  Hence if the equivalence principle is satisfied we can neglect gravity within the radius of curvature, $L >> \rho_0$.  This means that to leading order in $\frac{1}{L}$ the change in the energy of $\cal R$, conjugate to 
the Rindler time $\tau$ ,should be equal to the flux of the conserved energy current $J_a= T_{ab}\chi^b$ across the boundary, $\cal S$.   Hence this is a quantum mechanical expression of the equivalence principle.  

\item{} We call a quantum theory of gravity, {\it Unruh} if there is a  state\footnote{In the case that a theory is both Sciama and Unruh the two states $|R>$ should  be the same.  This can also be a mixed state.},  $|R >$, such that a thermometer at the timelike boundary $\cal S$ measures a temperature\cite{Unruh}
\f
T= \frac{\hbar a}{2\pi c}
\label{Unruh}
\ff
A thermometer can be modelled by a two state system dynamically coupled to the system $\cal R$ through an additional interaction term in the boundary Hamiltonian\cite{Eugenio1}.

\item{} We call a quantum theory of gravity {\it Bekenstein} if the change in entropy of the quantum system $\cal R$ is related to a change in 
area of $H$  by
\f
\delta S =\frac{\delta A_{H}}{4 G \hbar}
\label{Bekenstein}
\ff
Here by $\delta A_H$ we mean
\f
\delta A_{H} = {A}_{H^+ } (\lambda_+ ) -    <\Psi_- | \hat{A}_{H} (\lambda = 0)|\Psi_- >
\label{deltaAH}
\ff
where $\lambda^+$ is a time slice of $H^+$ which is after the support of $J_a$ on $H^+$.  We note that this is within the classical
region and hence ${A}_{H^+ } (\lambda_+ ) $ is a classical quantity.

\end{itemize}

\subsection{A quantum version of Jacobson's argument}

We now assume all three properties hold and derive the Einstein equations on 
 $H^+$.  To show this we follow Jacobson's argument, with slight modifications.  

We begin by discussing some properties of the energy momentum tensor.
We will assume that the matter perturbation $\delta T_{ab}$ has support on a finite region of $\cal S$ and is purely incoming, so that it vanishes of
$\Sigma_-,  W$ and $H$ as shown in Figure 2. By any finite $\tau^+$ the energy-momentum that flowed in through $\cal S$ is still contained in $\cal R$ and hence will
be measurable on $\Sigma^+$. 

We note that from the Sciama property (\ref{Ted1}) and 
(\ref{conserve2}) we have, again to leading order in $\frac{1}{L}$, 
\f
\delta E = \int_{\Sigma^+} J_a N^a  =  a \int_{H^+}   d^2 \sigma \sqrt{h} d\lambda \lambda T_{ab}k^a k^b
\label{DeltaE}
\ff
where $k^a=(\frac{ \partial }{ \partial \lambda } )^a$ is the null normal to the horizon.
For the second equality we use the conservation of $J_a$ in the classical region between $H^+$ and $\Sigma^+$.

We then apply the first law of thermodynamics
to the quantum system in $\cal R$, 
\f
\delta S =\frac{\delta E}{T}
\label{firstlaw}
\ff
By the Bekenstein condition and (\ref{DeltaE} ) this gives
\f
\delta S = \frac{\delta E}{T} = \frac{2\pi}{\hbar}  \int_{H^+}   d^2 \sigma \sqrt{h} d\lambda \lambda T_{ab}k^a k^b
\ff
But we can use the Raychaudhuri equations on $H^+$, because it is in the classical region.  This gives us
\f
\delta S = \frac{\delta A_H}{4G\hbar}  = \frac{1}{4G\hbar} \int_{H^+} d^2 \sigma \sqrt{h} d\lambda \lambda R_{ab} k^a k^b 
\ff
Jacobson then notices that this implies that on $\cal H^+$
\f
\left ( R_{ab} - 8\pi G T_{ab} \right ) k^a k^b =0
\label{Ted2}
\ff
but the starting point was the selection of the two surface $H$ which was arbitrary.  Hence (\ref{Ted1}) holds for every suitably defined null surface, which implies,
using the Bianchi identities, that the Einstein equations hold in $\cal M$,  
\f
R_{ab}-\frac{1}{2} g_{ab} R + \Lambda g_{ab} = 8\pi G T_{ab} .
\ff
for some value of the cosmological constant, $\Lambda$.  

Before we go on to loop quantum gravity there is one subtlety we need to understand. Jacobson's definition utilizes $\delta A_ H$, 
defined by (\ref{deltaAH}), which is
 the change in 
area on the horizon.  However the quantum theory of gravity will not tell us this directly; instead it will tell us $\delta A_{\cal S}$, which is the change in the area of the 
timelike surface $\cal S$.
This is defined by, 
\f
\delta A_{\cal S} = < \Psi_+ | \hat{A}_{\cal S} (\tau_+ )|\Psi_+ > -   < \Psi_- | \hat{A}_{\cal S} (\tau_-)|\Psi_- >
\ff
However it is straightforward to show that to arbitrary precision, if we make $\tau^+$ large we have,
\f
\delta A_{\cal S} = \delta A_{H}. 
\label{equalA}
\ff
First we note (\ref{initial}) so that the expectation values of the areas of $H$ and of ${\cal S}( \tau^- )$ are equal.  

We then need to show that the area of $H^+ (\lambda^+ )$ can be made arbitrarily close to the area
of ${\cal S}(\tau^+)$ by making $\tau^+$ large.  This  is a classical problem, because the two sphere denoted by ${\cal S}(\tau^+)$ is at a corner of the region $\cal R$ given by the intersection of the
outer timelike boundary $\cal S$ with the future space like surface $\Sigma_+$, while  $H^+ (\lambda^+ )$ is within the classical region.
So under the assumption that $a$ and $\tau^+$ are large and the curvature of the classical region is small we can use the Raychaudhuri equations to 
estimate the change in area on moving from the point $\lambda^+$ 
on the horizon to $\tau^+$ on the intersection of $\Sigma^+$ and $\cal S$.  
\f
\frac{\delta A}{A}= \frac{{A}_{H^+ } (\lambda_+ ) -  < \Psi_+ | \hat{A}_{\cal S} (\tau_+ )|\Psi_+ > }{{A}_{H^+ } (\lambda_+ )}
\ff
Choosing $\lambda^+$ so that it corresponds to the same $t$ in Minknowski coordinates as $\tau^+$ we find
\f
\left | \frac{\delta A}{A} \right | < \frac{c^2}{L^2 a^2} \ e^{-a \tau^+/c}
\label{error}
\ff
where  we recall that $L$ is a lower bound on the radius of curvature of the metric in the entire region.  But since 
$a$ is chosen so that $\rho_0= a^{-1} << L$ and  $\tau^+$ 
can be taken arbitrarily large this error can be made arbitrarily small. 

\section{Satisfying the criteria in loop quantum gravity}

We can now discuss the extent to which spin foam models satisfy the three criteria just stated.  

\subsection{The Bekenstein and Unruh properties}

The Unruh property (\ref{Unruh})  was established by Bianchi directly by coupling a thermometer carried by the accelerated detector to 
quantum spin network states in $\cal R$.  He also 
defines the quantum boost  Hamiltonian $\hat{H}$ (\ref{boostH}) and 
shows that on physical spin network states it satisfies 
\f
<\hat{H} > = \frac{a}{8\pi G} < \hat{A}_{\cal S} >  
\label{Bianchi2}
\ff
This implies that 
\f
\delta S =\frac{\delta A_{\cal S}}{4 G \hbar}
\label{quasiBekenstein}
\ff
However, making use of (\ref{equalA}) we see that 
(\ref{Bekenstein}) is satisfied so that the theory defined by a spin foam model is Bekenstein.

\subsection{The Sciama property and the FGP energy}

We showed above that the FGP energy is in fact the canonical energy arising from the boundary term of the
Holst action, when that boundary term is evaluated on the world line of the accelerating observer $\cal S$.  This allows us to identify it as the physical energy of the system $\cal R$ as seen by the accelerating observers on $\cal S$.

To show the Sciama property we note first that Bianchi shows that there is a state $|R>$ such that (\ref{Bianchi2}) holds.  Thus, the
expectation value of the boost hamiltonian (\ref{boostH}) is the FGP energy, which we  have shown is the canonical Hamiltonian.
This further supports the identification of the boost generator (\ref{boostH}) with the canonical hamiltonian for evolution in $\tau$.  

We next note that FGP use one component of the Einstein equations 
\f
0 = (G_{ab} -8 \pi G T_{ab}) k^a k^b
\label{Ein1}
\ff
together with (\ref{conserve2}) 
to show that
\f
\delta H_{\cal S} = \int_{\cal S} J_a N^a
\ff
thus establishing the Sciama property.

This of course required using a component of the Einstein equations.  However, this component is just proportional to a linear combination of constraints
and these are assumed to hold quantum mechanically in the quantum near horizon region $\cal R$.  
To see this we note that $k^a= \frac{1}{a \lambda } \chi^a $ in the limit
that we go to the horizon.  So we can write 
\f
0 =\int_{ H^+} \lambda (G_{ab} -8 \pi G T_{ab}) k^a k^b d\lambda dA =  \int_{ H^+} \ (G_{ab} -8 \pi G T_{ab})  \chi^a d\Sigma^b  
\label{Ein2}
\ff
using $d\Sigma^a = k^a d\lambda dA$ on $H^+$.  Using Gauss's law in the classical region bounded by $H^+$ and $\Sigma^+$, we  have 
\begin{eqnarray}
\int_{ H^+}  (G_{ab} - 8 \pi G T_{ab})  \chi^a d\Sigma^b &=& \int_{ \Sigma^+}  (G_{ab} -8 \pi G T_{ab})  \chi^a d\Sigma^b 
 \\
&=& \int_{ \Sigma^+}  (G_{ab} -8 \pi G T_{ab})  \chi^a N^b d\rho dA. \nonumber
\label{Ein3}
\end{eqnarray}
where on $\Sigma^+$ we use $d\Sigma^a = N^a d\rho dA$.  But $(G_{ab} -8 \pi G T_{ab})  \chi^a N^b $ is proportional to a linear combination 
 of the Hamiltonian and diffeomorphism constraints on $\Sigma^+$.    
 
We note that we have assumed that the state on $\Sigma^+$ satisfies the quantum constraints, (\ref{constraints}).  This implies that the
expectation value of the constraints are satisfied, 
\f
<\Psi (\tau^+ ) | \hat{\cal C} | \Psi (\tau^+ ) > =0.
\ff
But we also require that the classical fields on the boundary $\Sigma^+$ match the expectation values of quantum fields.  This implies that
the constraints as functions of the expectation values of quantum fields are satisfied up to terms of order $\hbar$ coming from operator ordering
and regularization.   

Spin foam models are constructed to propagate states which are solutions to the quantum constraints, because the spin foam amplitudes
give matrix elements, in the spin network basis, of the projection operator onto physical states.  Thus, (\ref{constraints}) hold in
spin foam models and we can then conclude that  (\ref{Ein1}) is satisfied on $H^+$ to leading orders in $\frac{1}{L}$ and $\hbar$.

Thus, there is good evidence that all three properties are satisfied by spin foam models.  This supports the conclusion that the Einstein equations hold in the course grained description when the quantum
dynamics in $\cal R$ is described by a spin foam model.  

\subsection{Consistency check}

Our derivation assumed that the geometry in a region of size $L >> \rho_0$ around the quantum near horizon region and $H^+$  can be taken to be approximately  flat, so that the radius of curvature is greater than $L$.  
But we have shown that there is curvature on $H^+$ due to the Einstein equations being satisfied with the source $\delta T_{ab}$.  For this to be consistent the curvature sourced by $\delta T_{ab}$ must have a radius of curvature greater than $L$.  This implies a bound on $\delta T_{ab}$, namely that
given by (\ref{bound}),

However, we would hope that the matter source could include at least one quanta whose wavelength fits inside of the near horizon region of the observer, otherwise it would not make sense to treat the source classically.  This requires that
\f
|\delta T_{ab}| > \frac{\hbar }{\rho_0^4}
\ff
This implies that the observer's acceleration and $L$ must be chosen so that
\f
\sqrt{l_p L} < \rho_0 < L
\ff
So the observer cannot hover within a planck scale of the  horizon. There must be room for a semiclassical region between the observer's world line and their horizon.   Note, however,  that (\ref{error}) can still be satisfied arbitrarily well by 
waiting to a time $\tau^+$ large in units of $\rho_0 / c$.

\section{The leading contribution to the spin foam calculation.}

I can also discuss the leading term in the spin foam sum to the amplitude $Z(\Psi_\pm, T_{ab},\tau_\pm, a)$. 

Following Bianchi , we define an initial  four simplex as follows.  Represent a facet of $H$ by a trianglular face, $F$.   Define the vertex $v_-$ corresponding to the event $\tau_-$ on an observers world line, $x^a (\tau_- )$
in $\cal S$.  (See Figure 3).  $F$ and $v_-$ together make a tetrahedron, $T_-$.  
This corresponds to the initial slice $\Sigma_-$.

The initial state, $|\Psi_- >$,  on $\Sigma_-$,  is defined on a dual four valent node dual to
$T_-$ with four edges dual to its faces.  

Now, pick a second event  $x^a (\tau_1)$ to the future of $x^a (\tau_-)$ on the world line.  Construct a
one to four move on $T_-$ and identify the new vertex so added with  $x^a (\tau_1)$.  $F$ and  $x^a (\tau)$ together make a new tetrahedron,
$T_1$ which corresponds to and is contained in the space like slice $\Sigma_1$ that includes $F$ and  $x^a (\tau_1)$.  

The five points,  $x^a (\tau_- ),  x^a (\tau_1 )$ and the three vertices of $F$ make a four simplex, $U_1$to which a spin foam model will provide an amplitude. One can choose the state on the boundary of this four simplex in the following way.  $F$ is dual to a spin, $j_F$ while the labels on the graphs dual to $T_-$ and $T_1$ represent the initial and final state.  Label the  faces to correspond to a choice of edge lengths for all the edges that join 
$x(\tau_-)$ or  $x^a (\tau_1)$ to the nodes of $F$ to be $\rho_0 = \frac{1}{a}$.  The choice of labels is also constrained by the requirement that
the timelike edge between $x^a (\tau_-)$ and $x^a (\tau_-)$ has a fixed length.  These fixed lengths are large in Planck units. 

\begin{figure}[h!]
\begin{center}
\includegraphics[width=.5 \textwidth]{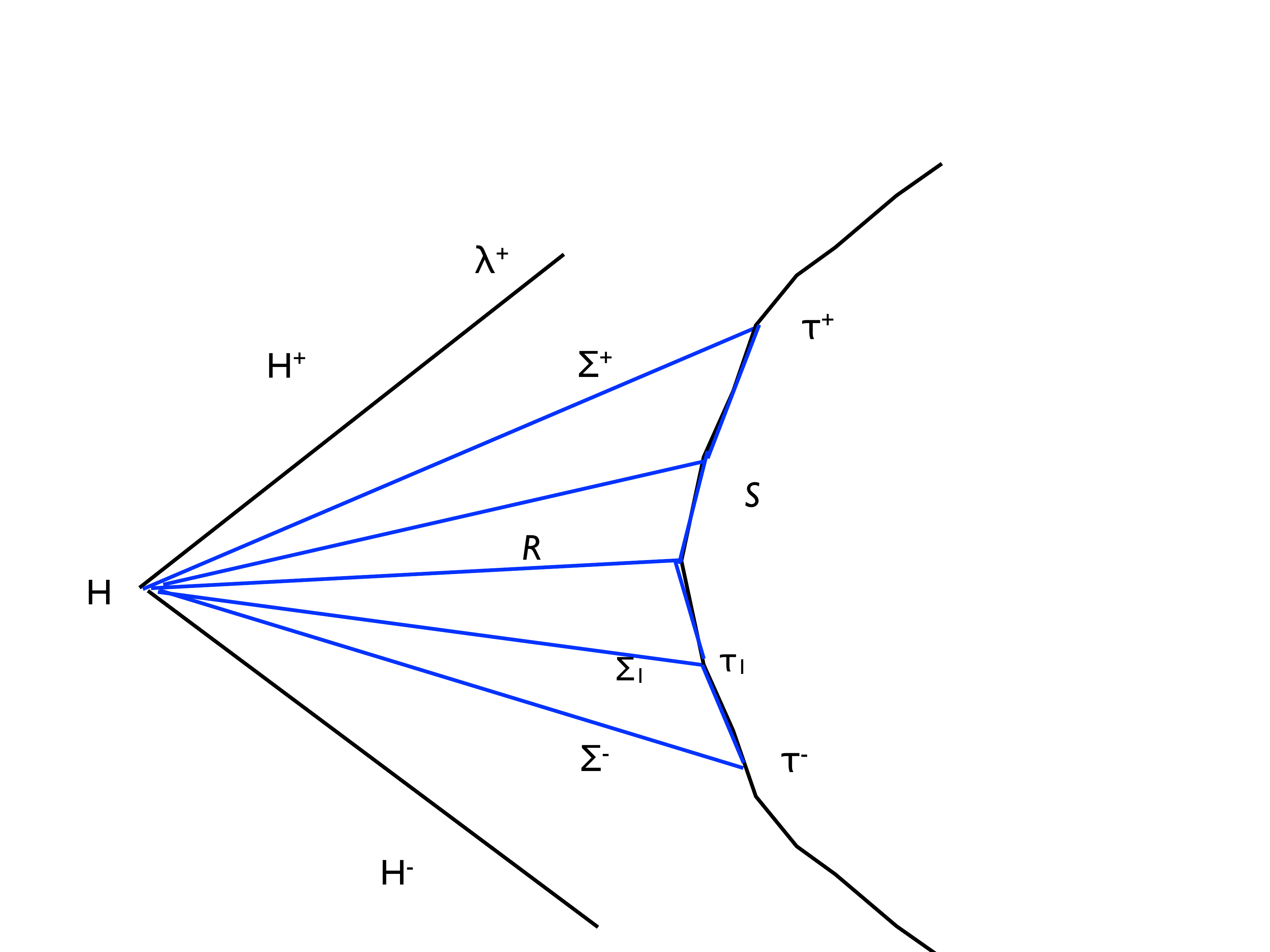}
\end{center}
\caption{The quantum near horizon region $\cal R$ constructed in a spin foam by a series of $1 \rightarrow 4$ moves, which are here illustrated in this two dimensional figure by $1 \rightarrow 2$ moves.}
\end{figure}
 Thus, we can write the four simplex amplitude as
\f
{\cal A}= <\Psi_1 | W(\Delta \tau , a ) |\Psi_- >
\ff
where $\Delta \tau = \tau_1-\tau_- $.

Now repeat the process.  Define a new point  $x^a (\tau_2)$ to the future of  $x^a (\tau_1)$ on the world line so that $\tau_2 =\tau_1 + \Delta \tau $ and identify it with the new vertex created by a one to four move operated on the tetrahedron $T_1$.  Repeat this $N$ times until  $\tau_N= \tau_+$.  The 
partition function is then
\f
Z=  < \Psi_+ | W^N (\Delta \tau , a ) |\Psi_- >
\ff

To complete the construction it is necessary to extend the triangulation transversely in the direction of the $x_\perp^i$ coordinates.  This can be done by following each $1 \rightarrow 4$ move with several $2 \rightarrow 3$ and $3 \rightarrow 2 $ moves to bring the triangulation of $\Sigma_1$ to the
same form as the initial triangulation of $\Sigma_-$.  I don't give the construction here.

\section{Conclusion}

These results strengthen the case that loop quantum gravity provides a physically plausible quantum theory of gravity, when its dynamics are expressed by spin foam amplitudes.  The argument assumes that a quantum space-time as described by a spin foam model has an emergent description in terms of a metric and energy momentum tensor on a manifold. Given that, the results discussed here give strong evidence that the Einstein equations govern that emergent description.

It is worth mentioning several issues that require more thought. 

\begin{itemize}

\item{} More clarification is needed concerning the role of the boundary conditions imposed at $\cal S$ and on $H$.  

\item{} There is more to understand about the detailed properties of the quantum states used in the derivations of the three properties of spin foam models.  

\item{} There is an alternative account  of black hole entropy in loop quantum gravity by Ghosh and Perez\cite{GP}, which makes use of a chemical potential to take into account the effects of thermodynamic processes that change the number of punctures, $N$,  on the boundary.  It is my understanding that these terms do not affect the calculations of Bianchi, but they do allow another setting in which the argument described here can be run in the context of a different thermodynamic ensemble in which variations of $N$ are physically meaningful\cite{Ale-personal}.

\item{} There is a non-relativistic argument for the emergence of Newtonian gravity as an entropic force given recently by Erik Verlinde\cite{Verlinde}.  This argument can also be run  in loop quantum gravity\cite{me-emergentgravity}.  It would be interesting to know how those arguments are related to the present one.
  
\item{} The present result does not preclude the existence of other routes from spin foam models to the conclusion that the Einstein equations are satisfied by the emergent geometry.  In particular, the fact that spin foam models yield states that satisfy the quantum constraints could be perhaps the basis of an alternate argument.  Nonetheless, note that even if we assume the existence of an emergent description labeled by coarse grained metric and energy momentum tensor, there need be no coupling between them at all, apart from covariant conservation. The metric might just be arbitrary and non-dynamcal.   Here it is the first law of thermodynamics, that requires the metric to have dynamics which is sourced by  the energy-momentum tensor.  Thus, even given that the thermodynamic argument is not the only route to the Einstein equations, it is a route that is illuminating.  Furthermore, it embraces and explains the basic lesson of black hole thermodynamics, which is that space-time itself is a thermodynamics system.  

\end{itemize}

These results are complementary to existing results concerning the low energy limit of spin foam models described in \cite{reviewSF}.  Those results make use of the asymptotic form of the spin foam amplitudes when the spins, or edge lengths, are taken to be large in Planck units\cite{asymptotic}.  There emerges in that limit a beautiful connection with classical general relativity in the form of Regge calculus.  The present discussion does not need to assume that spins or edge lengths are large, as these are not needed for the results in \cite{Eugenio1}).  However while the asymptotic results construct the space-time which emerges in the low energy limit through Regge calculus, the present results need to assume the emergence of the classical description.

At the same time, the present argument avoids having to discuss details of the ultraviolet behaviour of spin foam models such as whether one sums over triangulations or works with a fixed triangulation or whether summation is really equivalent to refinement.  This is because the argument is genuinely thermodynamic, as it makes use of the thermodynamic results of Bianchi.  The remarkable thing is that the quantum near horizon region generated by an accelerated observer is, according to the results of FGP and Bianchi, genuinely a thermodynamic system.  Unruh showed this was the case when the quantum system in question was an ordinary quantum field theory on Rindler space-time; what Bianchi shows is that this remains true when the quantum degrees of freedom are those of loop quantum gravity.  

These results may be extended to give corrections to the Einstein equations; indeed there is already a literature which describes how to use Jacobson's argument to reason from corrections to horizon entropy to corrections to the Einstein equations\cite{higher}.  

The present results may also  have implications for the issue of whether and how lorentz invariance is deformed in spin foam models, as Jacoson's thermodynamics is also a context for probing violations or deformations of Lorentz symmetry.  

In his groundbreaking paper, \cite{Ted1}, Jacobson argued that classical general relativity could emerge from a quantum statistical mechanics system that is not the quantization of classical general relativity.  This point is well taken, but neither is it excluded that the thermodynamic system the Einstein equations are emergent from would happen to be a quantization of general relativity.   Indeed, classical general relativity must be emergent from any successful quantum gravitational theory, as such a theory must be background independent and hence not describable in the language of classical fields on manifolds.  Jacobson's argument is then a central tool for exploring the connection between quantum and classical space-time geometry no matter whether the underlying quantum system is imagined or derived.  

\section*{ACKNOWLEDGEMENTS}

I am very grateful to Eugenio Bianchi and Alejandro Perez for patient explanations of their results.  I also would like to thank them and Andrzej Banbursky, Linqing Chen, Sabine Hossenfelder, Ted Jacobson and Carlo Rovelli for very helpful  comments on the manuscript.  The model of a horizon ini a causal spin foam presented in section $V$ was developed in unpublished work with Fotini Markopoulou.  Research at Perimeter Institute
for Theoretical Physics is supported in part by the Government of
Canada through NSERC and by the Province of Ontario through MRI.


\begin{thebibliography}{99}

\bibitem{Ted1}Ted Jacobson, {\it Thermodynamics of space-time: the Einstein equation of state,} Phys.Rev.Lett. 75 (1995) 1260-1263
e-Print: gr-qc/9504004.  

\bibitem{FGP}Ernesto Frodden, Amit Ghosh and Alejandro Perez, {\it A local first law for black hole thermodynamics}, arXiv:1110.4055.


\bibitem{Eugenio1} Eugenio Bianchi, {\it Entropy of non-extremal black holes from loop gravity}, arXiv:1204.5122.

\bibitem{reviewSF}For recent reviews of spin foam models please see Alejandro Perez, {\it The Spin Foam Approach to Quantum Gravity},
arXiv:1205.2019,  To appear in Living Reviews in Relativity; Carlo Rovelli, {\it Zakopane lectures on loop gravity  }, arXiv:1102.3660. 

\bibitem{CT}S. Carlip and C. Teitelboim, {\it The off-shell black hole,} Class. Quant. Grav.
12,1699 (1995).

\bibitem{MP}[42] S. Massar and R. Parentani, {\it How the change in horizon area drives black hole evaporation,} Nucl.Phys. B575, 333 (2000); gr-qc/9903027.

\bibitem{JP}Ted Jacobson, Renaud Parentani, {\it Horizon Entropy}, arXiv:gr-qc/0302099v1, 	Found.Phys. 33 (2003) 323-348.  

\bibitem{BW}Eugenio Bianchi and Wolfgang Wieland, {\it Horizon energy as the boost boundary term in general relativity and loop gravity}.

\bibitem{IMF}J.D. Bjorken, {\it  Asymptotic sum rules at infinite momentum}, 
Published in Phys.Rev. 179 (1969) 1547-1553; John B. Kogut* and Davison E. Soper,
{\it Quantum Electrodynamics in the Infinite-Momentum Frame},
Phys. Rev. D 1, 2901Ð2914 (1970).  

\bibitem{Unruh} W.G. Unruh, {\it Notes on black hole evaporation}, Phys. Rev. D14, 870 (1976).

\bibitem{Sciama}P. Candelas and D.W. Sciama, {\it Irreversible thermodynamics of black holes,}  Phys. Rev. Lett. 38, 1372 (1977):  
D.W. Sciama, {\it Black holes and fluctuations of quantum particles: an Einstein synthesis,}  in Relativity, Quanta, and Cosmology in the Development of the Scientific Thought of Albert Einstein, vol. II, M. Pantaleo, dir. and F. De Finis, ed. (Giunti Barb`era, Firenze, 1979).

\bibitem{Paddy}T.  Padmanabhan, {\it  Thermodynamical Aspects of Gravity: New insights},
arXiv:0911.5004,  Rep. Prog. Phys. 73 (2010) 046901;  {\it  Entropy of Static Spacetimes and Microscopic Density of States}, 
arXiv:gr-qc/0308070,  Class.Quant.Grav.21:4485-4494,2004;  {\it      Equipartition of energy in the horizon degrees of freedom and the emergence of gravity}, arXiv:0912.3165,  Mod.Phys.Lett.A25:1129-1136,2010.  

\bibitem{boundary}Viqar Husain and Seth Major, {\it  Gravity and BF theory defined in bounded regions,}  arXiv:gr-qc/9703043v2, Nucl.Phys. B500 (1997) 381-401; see also   Lee Smolin,  {\it   Linking Topological Quantum Field Theory and Nonperturbative Quantum Gravity},  arXiv:gr-qc/9505028, J.Math.Phys.36:6417-6455,1995;      {\it   A holographic formulation of quantum general relativity},  arXiv:hep-th/9808191, Phys.Rev. D61 (2000) 084007.

\bibitem{Verlinde} Erik P. Verlinde,   {\it  On the Origin of Gravity and the Laws of Newton },  arXiv:1001.0785, JHEP 1104:029,2011. 

\bibitem{me-emergentgravity}L. Smolin, {\it Newtonian gravity in loop quantum gravity },   arXiv:1001.3668. 
  

\bibitem{GP}A. Ghosh and A. Perez, {\it   Black hole entropy and isolated horizons thermodynamics    }
6. arXiv:1107.1320.  
   
\bibitem{Ale-personal}Alejandro Perez, personal communication.

\bibitem{asymptotic}Conrady, Florian, and Freidel, Laurent, {\it On the semiclassical limit of 4d spin foam models}, Phys. Rev., D78, 104023, (2008). [DOI], [arXiv:0809.2280 [gr-qc]]; Conrady, Florian, and Freidel, Laurent, ÒPath integral representation of spin foam models of 4d gravityÓ, Class. Quant. Grav., 25, 245010, (2008). [DOI], [arXiv:0806.4640 [gr-qc]]; Barrett, John W., Dowdall, Richard J., Fairbairn, Winston J., Gomes, Henrique, and Hell- mann, Frank, {\it Asymptotic analysis of the EPRL four-simplex amplitude}, J. Math. Phys., 50, 112504, (2009). [DOI], [arXiv:0902.1170 [gr-qc]];  Barrett, John W., Dowdall, Richard J., Fairbairn, Winston J., Hellmann, Frank, and Pereira, Roberto, {\it Lorentzian spin foam amplitudes: graphical calculus and asymptotics}, Class. Quant. Grav., 27, 165009, (2010). [DOI], [arXiv:0907.2440 [gr-qc]].  

\bibitem{higher} Raf Guedens, Ted Jacobson and Sudipta Sarkar, 
{\it  Horizon entropy and higher curvature equations of state }, Phys.Rev. D85 (2012) 064017
e-Print: arXiv:1112.6215 [gr-qc]

\end{thebibliography}
\end{document}